\documentclass[paper,12pt]{article}
\usepackage{cite}
\usepackage{amsfonts}
\usepackage{amsmath,amsthm}
\usepackage{amssymb}
\usepackage{comment}
\usepackage{graphicx,multirow,xcolor}
\usepackage{subcaption}
\usepackage{float}
\usepackage[unicode=true,pdfusetitle,
bookmarks=true,bookmarksnumbered=false,bookmarksopen=false,
breaklinks=false,pdfborder={0 0 0},backref=false,colorlinks=false]{hyperref}
\usepackage{enumitem}
\usepackage{placeins}
\input epsf
\newtheorem{theorem}{Theorem}
\newtheorem*{acknowledgement*}{Acknowledgements}

\newtheorem{remark}[theorem]{Remark}

\newcommand{\rossella}[1]{\textcolor{blue}{#1}}

\newcommand{\mattia}[1]{\textcolor{cyan}{#1}}
\definecolor{dgreen}{rgb}{0.078,0.418,0.184}
\newcommand{\sara}[1]{\textcolor{green}{#1}}

\title{A geometric analysis of the impact of large but finite switching rates on vaccination evolutionary games}
\author{Rossella Della Marca$^{1}$, Alberto d'Onofrio$^{2,*}$, Mattia Sensi$^{3,4,*}$, Sara Sottile$^5$\\[1em]
$^1${\footnotesize Mathematics Area, SISSA -- International School for Advanced Studies,}\\{\footnotesize via Bonomea 265, I-34136 Trieste, Italy}\\{\footnotesize rossella.dellamarca@sissa.it}
	\\[0.5em]
$^2${\footnotesize Department of Mathematics
		and Geosciences, University of Trieste, }\\{\footnotesize
	 via Alfonso Valerio 12/1, 34127 Trieste, Italy}\\{\footnotesize alberto.donofrio@units.it  (*corresponding author)}\\[0.5em]
$^3${\footnotesize MathNeuro Team, Inria at Universit\'e C\^ote d'Azur, 2004 Rte des Lucioles, 06410 Biot, France},\\
$^4${\footnotesize Politecnico di Torino, Corso Duca degli Abruzzi 24, 10129 Torino Italy}\\{\footnotesize mattia.sensi@polito.it  (*corresponding author)}\\[0.5em]
$^5${\footnotesize Department of Mathematics, University of Trento, Via Sommarive 14, 38123 Povo - Trento, Italy} \\{\footnotesize  sara.sottile@unitn.it}}

\date{}

\begin{document}

\maketitle

\begin{abstract}
In contemporary society, social networks accelerate decision dynamics causing a rapid switch of opinions in a number of fields, including the prevention of infectious diseases by means of vaccines. This means that opinion dynamics can nowadays be much faster than the spread of epidemics. Hence, we propose a Susceptible-Infectious-Removed epidemic model coupled with an evolutionary vaccination game embedding the public health system efforts to increase vaccine uptake. This results in a global system ``epidemic model + evolutionary game''. The epidemiological novelty of this work is that we assume that the switching to the strategy ``pro vaccine'' depends on the incidence of the disease. As a consequence of the above-mentioned accelerated decisions, the dynamics of the system acts on two different scales: a fast scale for the vaccine decisions and a slower scale for the spread of the disease. Another, and more methodological, element of novelty is that we apply Geometrical Singular Perturbation Theory (GSPT) to such a two-scale model and we then compare the geometric analysis with the Quasi-Steady-State Approximation (QSSA) approach, showing a criticality in the latter. 
Later, we apply the GSPT approach to the disease prevalence-based model already studied in (Della Marca and d'Onofrio, \textit{Comm Nonl Sci Num Sim}, 2021) via the QSSA approach by considering medium-large values of the strategy switching parameter.
\end{abstract}

\textbf{Keywords:} fast-slow system, behavioural epidemiology of infectious diseases, entry-exit function, vaccine hesitancy, mathematical epidemiology, geometric singular perturbation theory

\section{Introduction}\label{sec:intro}
The increasing spread of hesitancy and refusal of vaccines is a major challenge for global public health. This problem was originally born  in the field of  prevention of childhood diseases and of influenza but it has largely been observed also during the current pandemic of COVID-19 \cite{troiano2021}. Focusing on childhood diseases vaccines, we may say that one of major determinants of this problem is the phenomenon of ``pseudo-rational" exemption to vaccination. This phenomenon consists in the fact that parents both overweight real or supposed side effects of vaccines and underweight real disease-related risks \cite{mado,spva}. This causes the public health systems (PHSs) to spend considerable energies and budget to mitigate the impact of this phenomenon, typically by means of vaccine awareness public campaigns aimed at increasing the vaccine uptake. 

In consequence of  the outbreak of exemption to vaccination, a new scientific discipline has been developed: the Behavioural Epidemiology of Infectious Diseases (BEID) \cite{mado,spva}, whose aim is the inclusion in epidemic models of the description of human decision making (concerning, e.g., vaccination choices, social distancing, mobility patterns).

Not surprisingly, Game Theory has an important place in the context of BEID. Namely, the dynamics of vaccine decision is very frequently modelled as an imitation evolutionary game \cite{bauch05,domapo11} which can be represented  as an infection of ideas process \cite{spva}. Of course, also the modeling of the above-mentioned efforts of PHSs aimed at increasing the vaccine uptake have been introduced in the BEID literature  \cite{domapo}.

In the papers \cite{bauch05,domapo,domapo11} the implicit focus was on slow  changes of vaccine strategy. In the age of social media this implicit assumption is often unrealistic. Indeed, social networks accelerate decision dynamics causing a rapid switch of opinions in a number of fields, including the prevention of infectious diseases, for example in vaccination campaign. This means that opinion dynamics can nowadays be faster than the spread of epidemics. We are witnessing what has been defined as an {``exponential growth in public opinion channels"} \cite{buildOpinion}, which leads public opinion to be extremely volatile on many key subjects  \cite{alberici2013influence}, such as  politics \cite{alberici2013influence,Jost}  and vaccines \cite{bron,onparksong}. This scenario, unimaginable until few years ago,  must guide contemporary models in BEID.

Recently, Della Marca and d'Onofrio \cite{DELLAMARCA2021} explored the impact of the above-mentioned volatility on the modelling of public response to vaccine awareness campaigns for favouring vaccine uptake. Since the evolutionary vaccination game is endowed of a parameter that tunes the velocity of strategy change, under the hypothesis that such parameter is extremely large, they applied a Quasi-Steady-State  Approximation (QSSA) of the model \cite{domapo}. This resulted in a Susceptible--Infectious--Removed (SIR) epidemic model with a nonlinear dependence on the control: the PHS effort to increase the vaccine uptake. This control was designed via optimal control approach and numerically implemented also via heuristic global optimization methods.

In the present work, we drastically depart from the paper \cite{DELLAMARCA2021} in a number of modeling and methodological points. Firstly, here we hypothesize that the switching from the vaccine refusal to vaccine acceptance is influenced by the information on current disease \textit{incidence}. This is a more realistic hypothesis since the most widely diffused information on the spread of an infectious disease is not the prevalence but the incidence. Secondly, apart from the extreme cases that the velocity of strategy change is small or is practically infinite, we consider an intermediate case: the study of the impact of a large but finite switching velocity via the Geometric Singular Perturbation Theory (GSPT) \cite{fenichel1979geometric}.

GSPT \cite{fenichel1979geometric} is a powerful approach to model phenomena evolving on multiple time scales \cite{bertram2017multi,hek2010geometric,jones1995geometric,kuehn2015multiple}. Such separation in time scales differing by many orders of magnitude is quite common in real world scenarios (e.g., chemical oscillations, neuroscience \cite{desroches2008mixed,desroches2018spike,rodrigues2016time,sensi2023slow,taher2022bursting}, electromechanical devices \cite{em}, lasers, ecology, celestial mechanics, pattern formation \cite{kuehn2015multiple}). In particular, it has been applied to epidemics models in which immunity windows and demographic turnover are much longer than infectious periods \cite{aguiar2021time,castillo2016perspectives,jardon2021geometric,jardon2021geometric2}, and to discrete time epidemic models \cite{de2020discrete,bravo2021discrete,bravo2017discrete,de2017discrete}.

In this work, in line with what was done e.g. in the paper \cite{schecter2021},
we take into the account that the {time scale}
of vaccination strategy changes, although very {fast, nonetheless is not instantaneous}. Hence, we model the volatility of strategy switching by means of rigorous GSPT. We compare our geometric analysis to a classical QSSA, highlighting when the two approaches lead to the same conclusion, and when the QSSA instead fails {to reproduce the exact dynamics for a large but finite strategy switching rate}. This failure, as we will illustrate, is due to a delayed loss of stability of the critical manifold in our model. We exploit the so-called entry-exit function \cite{liu2000exchange,neishtadt1987persistence,neishtadt1988persistence,schecter2008exchange} to characterize this crucial part of the dynamics of our model.
Moreover, in the final part of this work we briefly apply the GSPT also to the model heuristically inferred in the paper \cite{domapo}.

The manuscript is organized as follows. In Section \ref{sec:background}, we introduce an evolutionary vaccination game where the information on the spread of the disease concerns the disease incidence; in Section \ref{sec:incidence}, we perform a qualitative analysis of the model in the case of low/medium strategy switching rate; in Section \ref{sec:QSSA}, we consider the case of very large (infinite) switching rate and apply the QSSA; in Section \ref{sec:GSPT_c2}, we investigate the case of  large but finite switching rate by using the GSPT approach and compare  the results to those obtained by the QSSA; in Section \ref{sec:numerics}, we provide some numerical simulations; in Section \ref{sec_app_C1}, we apply the  GSPT approach to the evolutionary vaccination game studied in the paper \cite{domapo}, where the information on the spread of the disease concerns the disease prevalence; we conclude in Section \ref{sec:concl}, summarizing our main results and providing inspiration for future research.

\section{Background of Evolutionary Vaccination Game as a process of Mutual Infection of ideas}\label{sec:background}
Let us consider the SIR-like model  describing the dynamics of a vaccine-preventable endemic childhood disease under voluntary vaccination choices \cite{domapo,domapo11}:
	$$\begin{aligned}
	\dot{S}&=\mu(1-p(\tau))-\mu S-\beta SI,\\
	\dot{I}&=\beta SI - (\nu+\mu) I ,
\end{aligned}$$
where: i) $\tau$ denotes the {(slow)} time variable; ii) $S$ and $I$ represent the fraction of susceptible and infectious individuals within the population at time $\tau$; iii) birth and natural mortality rates are equal to a value $\mu$; iv) $p$ is the time-dependent vaccine uptake of newborns; v) $\beta$ is the disease transmission rate; vi) $\nu$ is the recovery rate. 

Note that (iii) implies a stationary population, so that we can neglect the equation ruling the dynamics of the fraction of removed individuals: $R=1-S-I$.
Observe that, independently of the dynamics of $p$, the basic reproduction number   is
\begin{equation}
    \mathcal{R}_0=\frac{\beta}{\nu+\mu}.\label{R0}
\end{equation}
To model the impact of human decision making on the vaccination choices, we  assume that
the population of parents is proportional to the total
(constant) population and is divided into two groups  \cite{bauch05,domapo11,domapo}: ``pro-vaccine'' and ``anti-vaccine''. The first group is given by parents who are in favour of vaccines and vaccinate their children ($p$); the second one is given by parents who are hesitant or overtly against vaccination and, as a consequence, do not vaccinate their children ($a=1-p$). 

The evolution of $p$ follows an imitation game dynamics that could be inferred by employing an economics-oriented approach based on the concept of payoffs \cite{bauch05,domapo11} or, in alternative, a statistical physics-oriented approach \cite{BBCAFF,spva}. We follow the second approach that we consider to be much clearer than the first one in the present context. The basic concept is that the dynamics of $p$ and $a$ are ruled by a ``double contagion'' of ideas between the two involved groups. This approach yields the following family of models \cite{spva}: 
$$\begin{aligned}
\dot p&=k_1 \bar\theta(M_{d}) p a -k_1 \bar\alpha(M_{v}) a p,\\
\dot a&=-k_1 \bar\theta(M_{d}) p a +k_1 \bar\alpha(M_{v}) a p,
\end{aligned}  $$
where: i) the ``force of infection'' concerning the switch from the strategy ``anti-vaccine'' to the strategy ``pro-vaccine'' is $ k_1 \bar\theta(M_d) p $, where $M_{d}(\tau)$ is an information variable on the extent of the disease status in the community (e.g., incidence, prevalence); ii) the ``force of infection'' concerning the switch from the strategy ``pro-vaccine'' to the strategy ``anti-vaccine'' is $k_1 \bar\alpha(M_{v}) a $, where $M_{v}(\tau)$ is an information/``rumors'' variable on the extent of vaccine-related side effects; 
iii) the parameter $k_1$ is a time scale tuning parameter that characterizes the velocity of strategy switching in the population of parents. 
In the following, we assume that both $\bar\theta(\cdot)$ and $\bar\alpha(\cdot)$ are linear-affine functions, namely we set:
\begin{equation}\label{theta_alpha_gen}
\bar\theta(M_d)=\theta_0+\theta_1 M_d,\,\,\bar\alpha(M_{v})=\alpha_0+\alpha_1 M_{v},
\end{equation}
with $\theta_0\geq \alpha_0$.
The action of the PHS to favour the vaccine uptake is simply modeled as an additional switch, say $k_1 \gamma_1 a $, from the strategy ``anti-vaccine'' to the strategy ``pro-vaccine''. This provides the model:
$$\begin{aligned}
\dot p&=k_1 (\theta_0 +\theta_1 M_d)  p a -k_1(\alpha_0+\alpha_1 M_{v})  a p+ k_1 \gamma_1 a,\\
\dot a&=-k_1 (\theta_0 +\theta_1 M_d)  p a +k_1 (\alpha_0+\alpha_1 M_{v})  a p-k_1 \gamma_1 a.
\end{aligned}  $$
Taking into the account that $a=1-p$ yields the following imitation game equation for $p$:
\begin{equation}
		\dot{p}= k_1 p(1-p)(\theta_0 -\alpha_0+ \theta_1  M_d  -\alpha_1 M_v  )+ k _1\gamma_1 (1-p).\label{eqt_imit}
\end{equation}%
It is easy to see that one cannot identify all of these parameters:$$ (k_1,\alpha_1,\gamma_1). $$
The reason is simple: $k_1$ never appears ``alone'' so, unless one has an \textit{a priori} knowledge of the value of $\gamma_1$ or of $\alpha_1$, $k_1$ cannot be obtained from a parameter estimation of epidemiological data using our model. 
\begin{remark}
    The scenario is similar to the Malthusian parameters $b$ and $m$:
$$ \dot{x} = (b-m) x = r x. $$
Unless one has an external measure of $b$ or of $m$, one cannot identify  both $b$ and $m$ from data concerning $x(t)$. One can only fit the difference 
$$ r= b-m .$$
\end{remark}
Thus, the best option is to rewrite the imitation game equation (\ref{eqt_imit}) rescaling the parameters by $\alpha_1$, as follows 
\begin{equation}
    \label{normaliz}
 k= k_1 \alpha_1, \quad \delta= \frac{\theta_0-\alpha_0}{\alpha_1}, \quad \theta = \frac{\theta_1}{\alpha_1} ,  \quad \gamma = \frac{\gamma_1}{\alpha_1}.\end{equation}
As mentioned in  Section \ref{sec:intro}, we focus here on the relevant case in which the main information on the disease available to the population is the number of new cases. Thus, $M_d$ is a measure of the available information on the disease \textit{incidence} at time $\tau$. In the case in which the decisions are taken by only considering current information, we obtain 
\begin{equation}
	M_d(\tau)=\beta S(\tau)I(\tau).
\end{equation}
We further assume that the information on vaccine side-effects is proportional to the vaccine uptake of newborns, namely $M_v(\tau)=p(\tau)$, like in the papers \cite{domapo,domapo11,DELLAMARCA2021}.
We thus obtain the following complete model:
\begin{subequations}
	\begin{align}
		\dot{S}&=\mu(1-p)-\mu S-\beta SI,\\
		\dot{I}&=\beta SI - (\nu+\mu) I,\\ 
		\dot{p}&= k \left( p(1-p)\left(\delta + \theta\beta  SI  - p  \right)+ \gamma (1-p)\right).\label{dotpC2}
	\end{align}\label{SIRpcompleto}%
\end{subequations}

\section{An Evolutionary Game in which strategy switching depends on the incidence of the disease}\label{sec:incidence}

Let us start by investigating the case of low/medium rate of strategy switching  ($k\not\gg 1$), corresponding to classical societies with far less volatile opinions.

The model (\ref{SIRpcompleto}) presents two disease-free equilibria. The first one is a disease-free state in which all the newborns are vaccinated:
\begin{equation*}
    E^P = (0,0,1).
\end{equation*}
It is shown in Appendix \ref{App1} that high values of $\gamma$, i.e. $\gamma >\bar\gamma$, where $$\bar\gamma= 1 - \delta,$$ ensure the global attractivity of $E^P$. Conversely, when $\gamma < \bar\gamma$, $E^P$ is unstable.

The second disease-free equilibrium is $$E^0 = (1-p^0,0,p^0),$$
where 
\begin{equation}\label{p0}
p^0 = \dfrac{\delta + \sqrt{\delta^2 + 4 \gamma}}{2}, 
\end{equation}
which exists only when  $p^0<1$, namely $\gamma< \bar\gamma$. This disease-free state bifurcates from the equilibrium $E^0$ at $\gamma= \bar\gamma$.

Let us introduce the threshold value \begin{equation}\label{gamma_crit}
   \gamma_c=p_c( p_c - \delta)<\bar\gamma,
\end{equation}
with
$$p_c=1-\dfrac{1}{\mathcal{R}_0},$$
and $\mathcal{R}_0$ as given in (\ref{R0}). It is possible to show   that if $\gamma> \gamma_c$ (i.e. $p^0>p_c$), then $E^0$ is globally attractive (see Appendix \ref{App2}). Further, if $\gamma<\gamma_c$ (i.e. $p^0<p_c$), then $E^0$ is unstable. When $E^0$ becomes unstable, a unique (and epidemiologically meaningful) endemic equilibrium $$E^*= (S^*,I^*,p^*)$$ appears by a transcritical bifurcation at $\gamma=\gamma_c$. The components of $E^*$ read
\begin{equation}
\label{EEcomp}
    S^* = \dfrac{1}{\mathcal{R}_0}, \quad I^* = \dfrac{\mu }{\nu + \mu}\left( p_c - p^*\right),\quad
    p^* = \dfrac{\delta + {\theta\mu }p_c + \sqrt{\left(\delta +{\theta\mu }p_c\right)^2 + 4 \gamma \left(1 +{\theta \mu} \right)} }{2 \left(1 + {\theta \mu}\right)}.\end{equation} 
Straightforward calculations show that $I^*>0$ is equivalent to $p^0<p_c$, that is $\gamma<\gamma_c$.
The local stability of the equilibrium $E^*$ may vary with the model parameters, in particular Hopf bifurcations can occur, as we will prove shortly in Theorem \ref{thEEstability}. First, let us introduce, for simplicity, the notation
\begin{equation}\label{q123}
q_1=\mu+\beta I^*,\quad
q_2=(\nu+\mu)\beta I^*,\quad
q_3=\dfrac{\gamma}{p^*}+ p^*,
\end{equation}
with $I^*$ and $p^*$ as given in (\ref{EEcomp}).
\begin{theorem}\label{thEEstability}
The endemic equilibrium $E^*$ of model (\ref{SIRpcompleto}) is locally asymptotically stable (LAS) if $W\geq 0$, with
 $$W=q_1^2q_3+\theta\mu p^*\beta I^*(\beta I^*-\nu)+2\sqrt{q_1q_2q_3(q_1q_3+\theta\mu p^*\beta I^*)},$$
and $q_1,\,q_2,\,q_3$ as given in (\ref{q123}).
Otherwise, if $W<0$, then there exist two values $k_1,\,k_2$ with $0<k_1<k_2$, such that $E^*$ is unstable for $k\in(k_1,k_2)$, whereas it is LAS for $k<k_1$ or $k>k_2$. Hopf bifurcations occur at $k=k_i$, $i=1,2$. Moreover, in such a case, if $k \in (k_1,k_2)$ then the orbits are oscillatory in the sense of Yakubovich.
\end{theorem}
\proof 
The Jacobian matrix of system (\ref{SIRpcompleto}) evaluated at the equilibrium $E^*$ reads
$$J=\left(\begin{array}{ccc}
     -q_1& -(\nu+\mu) &-\mu  \\
     \beta I^* & 0 & 0\\
    kp^*(1-p^*)\theta \beta I^* & kp^*(1-p^*)\theta (\nu+\mu) & -k(1-p^*)q_3
\end{array}\right),$$
leading to  the characteristic polynomial
$$p(\lambda)=\lambda^3+a_1\lambda^2+a_2\lambda+a_3,$$
with
\begin{equation*}
    a_1=k(1-p^*)q_3+q_1>0,\quad
    a_2=k(1-p^*)(q_1q_3+\theta \mu p^*\beta I^* )+q_2>0,\quad
    a_3=k(1-p^*)q_2(q_3+\theta \mu p^* )>0,
\end{equation*}
where $q_1,\,q_2,\,q_3$ are given in (\ref{q123}).
The positiveness of the coefficients of $p(\lambda)$ rules out, by Descartes rule of signs, the
possibility of real positive eigenvalues, so that stability losses of the endemic state can only occur via
Hopf bifurcations. More precisely,
according to Routh-Hurwitz theorem, $E^*$ is LAS if and only if $a_1a_2-a_3$ is positive, equivalently
written as
$$f(k)=b_0k^2+b_1k+b_2>0,$$
with
\begin{equation*}
    b_0=(1-p^*)^2q_3(q_1q_3+\theta\mu p^*\beta I^*)>0,\quad
    b_1=(1-p^*)(q_1^2q_3+\theta\mu p^*\beta I^*(\beta I^*-\nu)),\quad
    b_2=q_1q_2>0.
\end{equation*}
We chose the strategy switching rate, $k$, as bifurcation parameter because it affects the stability
but not the existence of $E^*$.

Thus, if $b_1\geq 0$ and/or if $\Delta=b_1^2-4b_0b_2\leq 0$, then $f(k)>0$  independently of $k$. 

Otherwise, if $b_1<0$ and $\Delta>0$, then $f(k)$ has two positive roots $k_1,\,k_2$, with $0<k_1<k_2$, such that $f(k)<0$ for $k\in (k_1,k_2)$ and $f(k)> 0$ for $k<k_1$ or $k>k_2$.
In such a case, at $k_i$, $i=1,2$, the test for non-zero speed is fulfilled:
$$f'(k)|_{k=k_i}=\pm \sqrt{\Delta}\neq 0.$$
Note that, by simple algebra, one can write
$$\Delta=(b_1-2\sqrt{b_0b_2})(b_1+2\sqrt{b_0b_2}).$$
It follows that, if 
$$b_1+2\sqrt{b_0b_2}<0,$$
then $b_1<0$ and $\Delta>0$. 
Vice versa, if 
$$b_1+2\sqrt{b_0b_2}\geq 0,$$
then $b_1\geq0$ and/or $\Delta\leq 0$.

Finally, as regards the Yakubovich oscillatority, it holds that: i) the orbits of the system are bounded; ii)
the endemic equilibrium $E^*$ is unstable when $f(k)<0$; iii) the disease-free equilibria are unstable. Thus, we may apply the Yakubovich theorem \cite{EFIMOV2008,efimov09}. The claim follows.
\endproof
\begin{figure}[t!]
     \centering
         \centering
\includegraphics[width=0.6\textwidth]{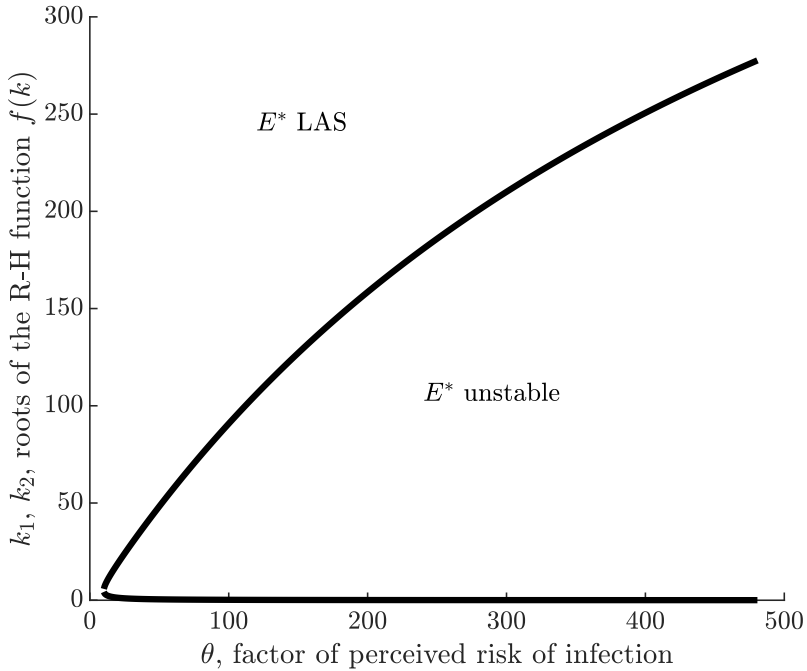}
     \caption{Hopf bifurcation locus for the equilibrium $E^*$  of model (\ref{SIRpcompleto}). Roots  $k_1,\,k_2$ of the Routh-Hurwitz function $f(k)$ in Theorem \ref{thEEstability} as functions of the factor of perceived risk of infection $\theta\in[0,25000/(\nu+\mu)]$. Parameter values as given in Table \ref{tab:param_table}, with $\psi=0.3$.}
     \label{figHopf}
\end{figure}
A bifurcation diagram showing the Hopf bifurcation locus $k_1,\,k_2$  as functions of the factor of perceived risk of infection $\theta$ is given in Fig. \ref{figHopf}. 

Regarding the Yakubovich oscillatority, this intuitively means that for sufficiently large time all the state variables are permanently oscillating, with regular or irregular
oscillations (periodic, quasi-periodic or chaotic). Formally, it holds that \cite{EFIMOV2008,efimov09}
$$-\infty<\liminf_{\tau\to +\infty} X(\tau)<\limsup_{\tau\to +\infty} X(\tau)<+\infty,$$
for $X\in\{S,I,p\}.$
Note that this is a global
result, unlike the Hopf bifurcation theorem, which is local.

\section{The case of very large $k$: Quasi-Steady-State Approximation}\label{sec:QSSA}

Let us investigate now the case of very large  rate of strategy switching, corresponding to the extreme case that the velocity of switching is practically infinite.

We consider system (\ref{SIRpcompleto}) with 
$$k =\frac{1}{\varepsilon},$$ 
and $\varepsilon>0$ very small.
The ensuing system is a slow-fast system   with two slow variables,
$S$ and $I$, and one fast variable, $p$.

If we assume that $\varepsilon\ll 1$, then a QSSA  for $p$   can be used, which yields
\begin{equation*}
	 p(1-p)(\delta+\theta\beta SI - p)+\gamma(1-p)=\varepsilon{\dot{p}}\approx 0.
\end{equation*}
As a consequence, in the limit $\varepsilon\rightarrow 0$, $p$ is the solution of the following algebraic equation:
\begin{equation}\label{eqQSSA}
0=(1-p)(\gamma+(\delta+\theta\beta SI)p - p^2),
\end{equation}
to be solved under the constraint $ 0 \leq p \leq 1$.
Equation (\ref{eqQSSA}) has two solutions: $p=1$, and the unique positive solution  of 
$$ \gamma+(\delta+\theta\beta SI)p - p^2=0. $$ 
Summarizing, $p$ tends to
\begin{equation}\label{bar p}
 p=\zeta(S,I)= \min\left(1,\dfrac{\delta+\theta\beta SI+\sqrt{(\delta+\theta\beta SI)^2+4\gamma}}{2}\right) 
\end{equation}
because it is stable, contrary to $p=1$  which is unstable. In other words: 
\begin{itemize}
    \item[i)] if $p=\zeta$, then $\dot p=0$ $\forall \varepsilon>0$; 
    \item[ii)] if $p \in [0, \zeta)$, then 
$$\lim_{\varepsilon \rightarrow 0}\dot p \rightarrow + \infty;$$
\item[iii)] if  
$  p \in (\zeta,1)$, then 
$$ \lim_{\varepsilon \rightarrow 0} \dot p \rightarrow - \infty.$$ 
\end{itemize}
Thus, model (\ref{SIRpcompleto}) reduces to the following  bidimensional model:
\begin{equation}\label{TPB}
\begin{aligned}
\dot{S}&=\mu(1-\zeta(S,I))-\mu S-\beta SI,\\
\dot{I}&=\beta SI - (\nu+\mu) I, 
\end{aligned}
\end{equation}
with $\zeta$ defined in (\ref{bar p}).


Model (\ref{TPB}) admits an unique disease-free equilibrium $$E^0=(1-p^0,0),$$
with $ p^0 = \zeta(S,0)$. When $\gamma<\gamma_c$, with $\gamma_c$ as given in (\ref{gamma_crit}), it also admits the endemic equilibrium $$E^*=\left(\dfrac{1}{\mathcal{R}_0},I^*\right),$$ where $I^*$ is the unique positive solution of
\begin{equation}
    \zeta\left(\dfrac{1}{\mathcal{R}_0},I^*\right)=p_c-\dfrac{\nu+\mu}{\mu}I^*.\label{inters}
\end{equation}
Indeed, the l.h.s. of (\ref{inters}) is an increasing function of $I^*$, the r.h.s. of (\ref{inters}) is a decreasing function of $I^*$, and $\zeta(1/\mathcal{R}_0,0)=p^0<p_c$. Note that, with a slight abuse of notation, we denote by $E^0$ and $E^*$ also the corresponding  equilibria of the model (\ref{SIRpcompleto}) in the case $\varepsilon\not\ll1$ (see Section \ref{sec:incidence}).

By using arguments similar to those of paper \cite{domasa}, it can be shown that, if 
$\gamma>\gamma_c$ 
then $E^0$ is globally asymptotically stable (GAS), see Appendix \ref{App3}; instead,  if $\gamma<\gamma_c$ then $E^0$ is unstable and $E^*$ is GAS in the positively invariant region $$\Omega^*=\{(S,I)|\, S\geq 0, \,I>0,\,S+I\leq 1,\,S\leq 1-p^0\}$$
(see Appendix \ref{App4}).

The proposed QSSA provides a model of the spread and control of an SIR-like infectious disease that extends the one in the paper \cite{domasa}  to the important case where the information on the disease spread is the  incidence, not the prevalence. As in the paper \cite{domasa}, oscillations do not occur since the rate of strategy change is too large. The physical reason of this lack of limit cycles and other oscillating structures is that here we are in a regime of extremely volatile public opinion. In other words, there is no opinion-induced delay with respect to the information on the disease spread.

\begin{remark}
The above procedure can easily be adapted to a far more general case in which $\bar{\theta}(\cdot)$ and $\bar{\alpha}(\cdot)$ are nonlinear, in place of (\ref{theta_alpha_gen}). This would result in a more general nonlinear relationship between $p(t)$, the instantaneous incidence ($\beta S I$) and the control parameter $\gamma$: $p=F(\beta S I; \gamma) \in [0,1]$.    
\end{remark} 

\section{An approach based on Geometric Singular Perturbation Theory}\label{sec:GSPT_c2}

In this section, we focus on a different approach to the model under analysis, through the use of techniques from GSPT. 

We denote by $\tau$ the slow time variable and with $t=k\tau=\tau/\varepsilon$ the fast time variable. 
In a more classical GSPT notation, system \eqref{SIRpcompleto} can be rewritten, in the fast time scale $t$, as
\begin{equation}\label{fast_ts}
	\begin{aligned}
		S'&=\varepsilon(\mu(1-p)-\mu S-\beta SI),\\
		I'&=\varepsilon(\beta SI - (\nu+\mu) I),\\ 
		p'&=  p(1-p)\left(\delta + \theta\beta  SI  - p \right)+  \gamma (1-p).
	\end{aligned}
\end{equation}
{Note that we denote by $\dot{X}$ the derivative of the variable $X$ with respect to $\tau$ and by $X'$ the derivative with respect to $t$, where $X\in \{S, I, p\}$.}
\begin{remark}\label{change_time}
    The unit of measure of $k$ is $1/$time. This implies that, when we apply a change in the time coordinate, bringing system \eqref{SIRpcompleto} to system \eqref{fast_ts}, the resulting $t = k  \tau=\tau / \varepsilon$ is dimensionless. Numerical simulations of \eqref{fast_ts} should be handled carefully, since $t$ does not have a time dimension; thus, we need to rescale the time accordingly.
\end{remark}
{The results obtained with GSPT are asymptotic as $\varepsilon \to 0$. Specifically, this means that for each result there exists a $\varepsilon_0>0$, which is often not explicitly quantifiable, such that the result holds for $0<\varepsilon<\varepsilon_0$. This translates, considering $k=1/\varepsilon$, to a validity of our results for $k \in (1/\varepsilon_0,+\infty)$.}

In some cases, the QSSA is in perfect agreement with the results obtained with GSPT. In other cases, namely for orbits which pass exponentially close in $\varepsilon$
to the manifold $\{ p=1 \}$, the QSSA is not able to replicate the delayed loss of stability of that same manifold. This behaviour is clearly visible in Fig. \ref{fig:vaccine}, in which different situations have been explored.

The critical manifold is given by
\begin{equation}\label{crit_manifC2}
	\begin{aligned}
    \mathcal{C}_0 =& \left\lbrace S\ge 0, I \ge 0, S+I \leq 1, p \in [0,1]\; |\; p(1-p)(\delta + \theta \beta S I - p) + \gamma (1-p) = 0 \right\rbrace\\
    = & \{p = 1\} \cup \{ p^2 - (\delta + \theta \beta SI)p - \gamma  = 0\} =: A \cup B .	\end{aligned}
\end{equation}
We first make the expression of $B$ explicit. To remain in the biologically feasible region, we impose $p\in [0,1]$, from which it follows that $p=\zeta(S,I) $, as given in \eqref{bar p}.
Also, it must be
\begin{align*}
   S I &\leq  \dfrac{1 - \delta - \gamma}{\theta\beta},
\end{align*}
where we have assumed that $2 - \delta - \theta \beta SI \geq 0$ on an open interval in $(0,1)^2$. Moreover, we need that $1 - \delta   > \gamma $, mirroring the necessary conditions coming from the paper \cite{domapo}.

Computing the Jacobian of the fast system \eqref{fast_ts} with $\varepsilon = 0$, we notice that there are two zero eigenvalues, corresponding to the slow variables $S$ and $I$. The third eigenvalue is 
\begin{align*}
    \lambda &= ( p^2 - (\delta + \theta \beta S I )p - \gamma ) + (1-p) (\delta + \theta \beta S I - 2  p).
\end{align*}
On $B$, the eigenvalue is
\begin{equation*}
    \lambda = (1-p) (\delta + \theta \beta SI - 2 p).
\end{equation*}
Note that $1-p \geq 0 $ and, on $B$, $ \delta
+\theta\beta S I - 2 p = -\sqrt{(\delta + \theta \beta S I)^2 + 4  \gamma} < 0$, thus $\lambda <0$ and $B$ is always locally attractive when it exists, independently of the value of $\mathcal{R}_0$.
 
Let us now focus on the behaviour of the system when $p\approx 1$ (i.e. close to $A$). If we sum the first two equations in $\eqref{SIRpcompleto}$, we obtain
\begin{equation*}
    \dot S+ \dot I \leq -\mu (S+ I), 
\end{equation*}
thus $S+I$ converges to zero. We remark that on $A$ the susceptible population $S$ can only decrease, since this set represents the situation in which there are no vaccine sceptical people, whereas $I$ is not always decreasing, although the sum $S+I$ is. The eigenvalue on $A$ is
\begin{equation*}
    \lambda =  p^2 - (\delta + \theta\beta SI)p - \gamma.
\end{equation*}
Thus,
\begin{equation*}
\begin{cases}
\lambda >0 &\text{if }\,  \dfrac{\delta + \theta\beta S I + \sqrt{(\delta + \theta\beta S I)^2 + 4 \gamma}}{2}<p\leq 1,\\
\lambda <0 &\text{if }\, 0 \leq p < \dfrac{\delta + \theta\beta S I + \sqrt{(\delta + \theta\beta S I)^2 + 4 \gamma}}{2}.
\end{cases}
\end{equation*}
Since the corresponding eigenvalue $\lambda$ changes its sign, the region $A$ is attractive until the intersection with the curve (\ref{bar p}), it then becomes repelling and, after a delay, the dynamics lands on the curve $B$. Depending on the value of $\mathcal{R}_0$ and $S_{\text{in}}$, where $S_{\text{in}}$ indicates the entrance of an orbit in a neighbourhood of $\{ p=1 \}$, we can observe two different behaviours:
\begin{itemize}
    \item Case I: if $\mathcal{R}_0 < 1$, or $\mathcal{R}_0 > 1$ and $S_{\text{in}}<1/\mathcal{R}_0$, then $I(\tau)$  is decreasing and the orbits converge to $B$ after a delay $T_E$ determined with the entry-exit function 
    $\int_0^{T_E}\lambda(\tau)\text{d}\tau=0$; hence the exit time $T_E$ is given implicitly by
    $$
    (1 -\delta - \gamma )T_E= \theta\beta\int_0^{T_E}S(\tau)I(\tau)\text{d}\tau.
    $$  
    A similar entry-exit phenomenon, with orbits eventually landing on a different branch of the critical manifold, was already observed  in the paper \cite{achterberg2022minimal}. We note that, however, we have no explicit formula for $S(\tau)I(\tau)$ on $p=1$. There is, to the best of the authors' knowledge, no theoretical result that justifies the canard-like behaviour of orbits remaining in a neighbourhood of an unstable branch in this setting. However, from the lower dimensional case of Section \ref{sec_app_C1}, for which the delayed loss of stability is theoretically foreseeable, and from our simulations, we conjecture that this is always the case, for the parameter values we are interested in;
    \item Case II: if $\mathcal{R}_0 >1$ and $S_{\text{in}}>1/\mathcal{R}_0$, then $I(\tau)$ changes its monotonicity and the convergence towards $B$ only happens after a short excursion away from it. In this case, the known formulas for entry-exit functions cannot be applied.
\end{itemize}

\section{Numerical simulations: the impact of finite $k$}\label{sec:numerics}
\begin{table}[t]
    \centering
\begin{tabular}{|c|c|c|}
\hline
\textbf{\textbf{Parameter}} &\textbf{\textbf{Formula}}    &\textbf{\textbf{Value}}          \\ \hline
$\nu$                               & -         & 52 years$^{-1}$                                   \\ \hline
$\mu$                                 & -       & 1/78 years$^{-1}$                                 \\ \hline
$\varepsilon$                         &  $0.1/\nu$        & $ 0.0019$ years  
  \\ \hline
  $k$                         &  $1/\varepsilon$        & $ 520$ years$^{-1}$
  \\ \hline
$\beta$ & $\mathcal{R}_0(\mu+\nu)$& $936.23$ years$^{-1}$\\ \hline
$p_{c}$                            & $1-1/\mathcal{R}_0 $ & $0.944$ 
\\ \hline
$\mathcal{R}_0$                              & - & $18$          
\\ \hline
$\psi$                                       &- &$\{0.3; 0.8\}$                                    \\ \hline
$\delta$                                     & $\psi p_{c} $&$ \{0.283; 0.756\}$   
\\ \hline
$\theta_{\text{pre}}$                        & -& $450$                                             \\ \hline
$\theta_{\text{bas}}$                        & ${\theta_{\text{pre}}}/{(\mu+\nu)}$ &$ 8.652$ years
\\ \hline
$\phi$                                       & -& $\{10 ; 50 \}$                                    \\ \hline
$\theta$                                     & $\phi\theta_{\text{bas}}$& $ \{86.52;432.59\}$ years 
\\ \hline
$\gamma_{c}$                       & $p_{c}(p_{c}-\delta)$ &$ \{ 0.624;0.178   \}$    \\ \hline
$\gamma$                                     & $\gamma_{c}/2$ & $\{0.089;0.312\}$  
\\ \hline
$S(0)$ &  ${1}/{\mathcal{R}_0}$ & $0.056$\\ \hline
$I(0)$ & ${\mu}\left(p_c - p(0) \right) /{(\mu + \nu)}$ & $4.793 \cdot 10^{-5}$
 \\ \hline
 $p(0)$& -&0.75
  \\ \hline
\end{tabular}
    \caption{Parameters values and initial conditions used for numerical simulations of models (\ref{TPB}) and (\ref{fast_ts}). }
    \label{tab:param_table}
\end{table}

For numerical simulations, we use epidemiological parameters compatible with a vaccine-preventable endemic childhood disease \cite{DELLAMARCA2021,domapo,domapo11}. In particular, we keep fixed the parameters values: $\nu=52$ years$^{-1}$, $\mu=1/78$ years$^{-1}$, $\mathcal{R}_0 = 18$, $\varepsilon = 0.1/\nu$.  At variance, we vary the values of the imitation game parameters   $\delta$, $\gamma$ and $\theta$, that are chosen in the following way:
\begin{itemize}
    \item since the value of $\delta$ affects the value of $\gamma_{c}$, recall \eqref{gamma_crit}, we define it as $\delta = \psi p_{c}$, where $\psi \in \{0.3,0.8\}$
    ;
    \item we then compute the value of $\gamma_{c} $ and we set $\gamma = \gamma_{c}/2$
    ;
    \item the most complex case concerns finding the value of $\theta$. We cannot use the same value used in the paper \cite{DELLAMARCA2021}, say $\theta_{\text{pre}}=450$, since in that case the perceived risk of infection is proportional to $I(\tau)$, whereas here it is proportional to $\beta S(\tau) I(t) $. We impose that, if $S(\tau)$ is close to its endemic equilibrium value $S^* = 1/\mathcal{R}_0$ and the value of $I(\tau)$ is equal to that in the prevalence-based case, then the two ``risks of infection" must be equal each other, namely  $ \theta \beta S^* I(\tau) = \theta_{\text{pre}} I(\tau)$. This gives in turn  a baseline value for $\theta$ in the present case, say $\theta_{\text{bas}} := \theta_{\text{pre}}/(\mu+\nu) $. We eventually set $\theta =\phi\theta_{\text{bas}}$, where $\phi \in \{10,50\}$.
\end{itemize}
As regards the initial conditions of the state variables, we assume that they are at the endemic equilibrium of the model (\ref{SIRpcompleto}) with constant vaccination $p(\tau)\equiv 0.75$; we choose this value because it is large enough, and it is representative of an initial condition in which three quarters of parents used to vaccinate their children. Therefore, 
 $S(0) ={1}/{\mathcal{R}_0}$, $I(0) = {\mu}\left(p_c - p(0) \right) /{(\mu + \nu)}$, $p(0)=0.75$.

The above-mentioned values of the parameters and the initial conditions are summarized in Table \ref{tab:param_table}.

Since $I(\tau)$ becomes exponentially small in $\varepsilon$, meaning $I=e^{-K/\varepsilon}$ for some $K>0$,
we apply a change of variables to system \eqref{fast_ts}, as reported in Appendix \ref{change_var}, in order to reduce its numerical stiffness. Moreover, recall from Remark \ref{change_time} that the time should be rescaled.

\begin{figure}[t!]
  \centering
\includegraphics[width=0.8\textwidth]{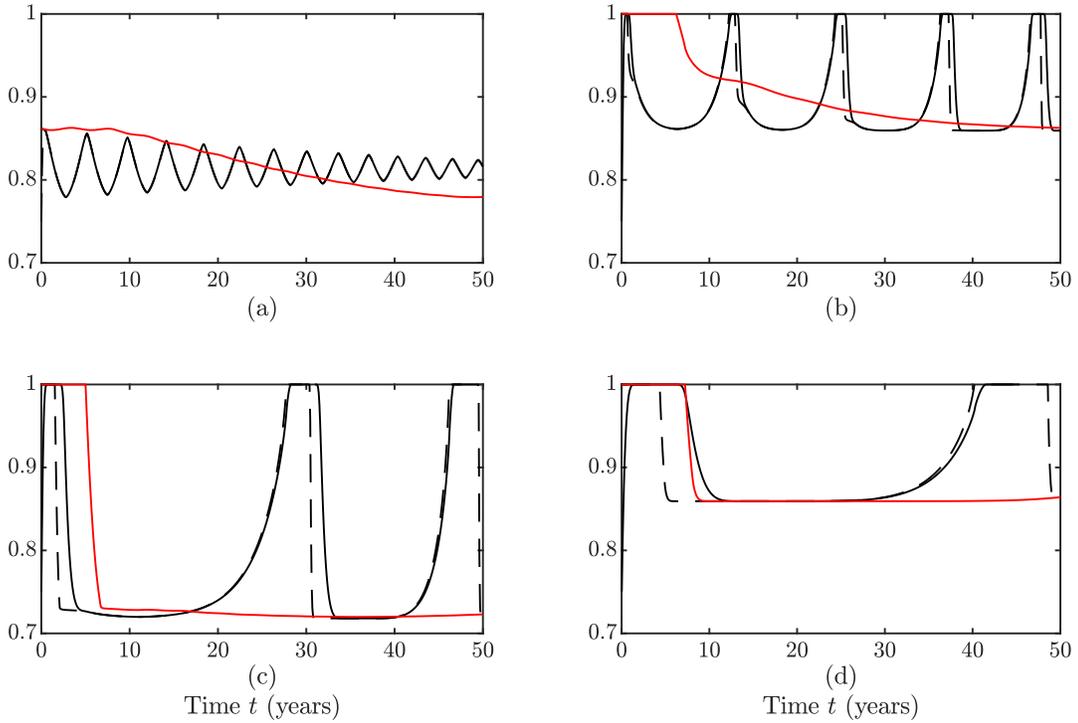}
    \caption{Impact of the finiteness of the rate of  switching strategy $k$. Vaccine uptake of newborns as predicted by the QSSA model (\ref{TPB}) (red lines) and by the GSPT model (\ref{fast_ts}) (black lines). The black dashed lines represent the critical manifold \eqref{crit_manifC2}. Panel (a):  $\psi=0.3,\,\phi=10$. Panel (b): $\psi=0.8,\,\phi=10$. Panel (c): $\psi=0.3,\,\phi=50$. Panel (d): $\psi=0.8,\,\phi=50$. Other parameter values and initial conditions as given in Table \ref{tab:param_table}.}
         \label{fig:vaccine}
     \end{figure}

\begin{figure}[t!]
 \centering
\includegraphics[width=0.8\textwidth]{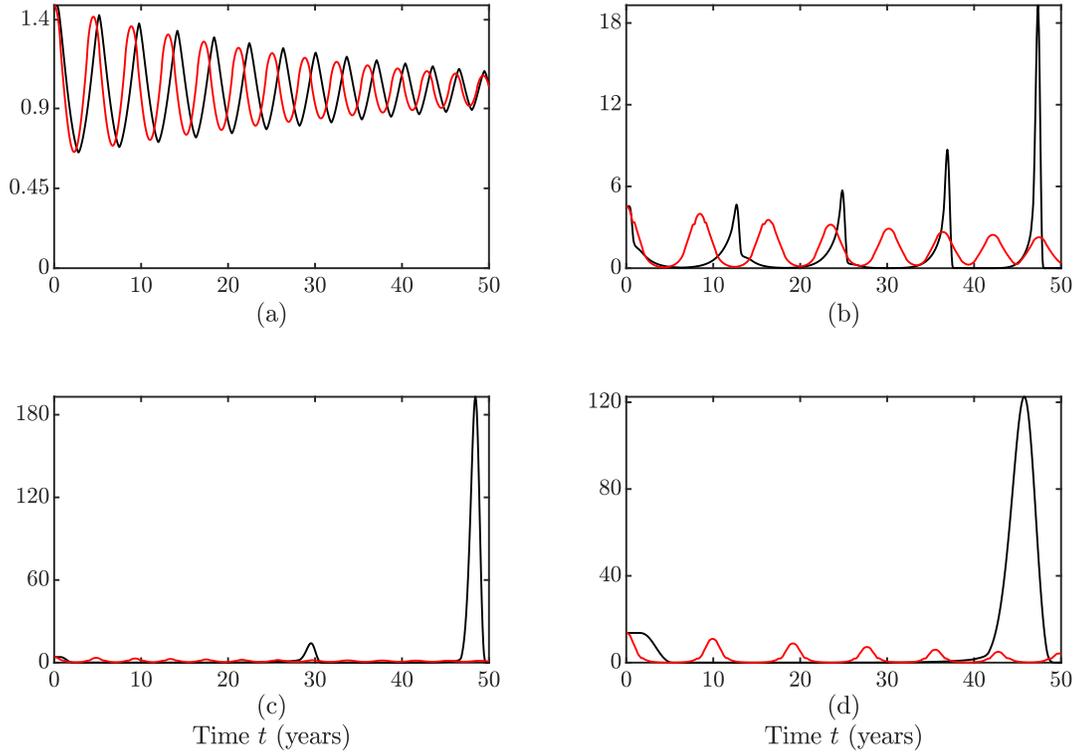}
    \caption{Dynamics of the fraction of infectious individuals as predicted by the QSSA model (\ref{TPB}) (red lines) and by the GSPT model (\ref{fast_ts}) (black lines), and normalized with respect to the endemic equilibrium value $I^*$, given in (\ref{EEcomp}). Panel (a):  $\psi=0.3,\,\phi=10$. Panel (b): $\psi=0.8,\,\phi=10$. Panel (c): $\psi=0.3,\,\phi=50$. Panel (d): $\psi=0.8,\,\phi=50$. Other parameter values and initial conditions as given in Table \ref{tab:param_table}.}
         \label{fig:dyn_epi_inf}
\end{figure}

\begin{figure}[t!]
 \centering
\includegraphics[width=0.8\textwidth]{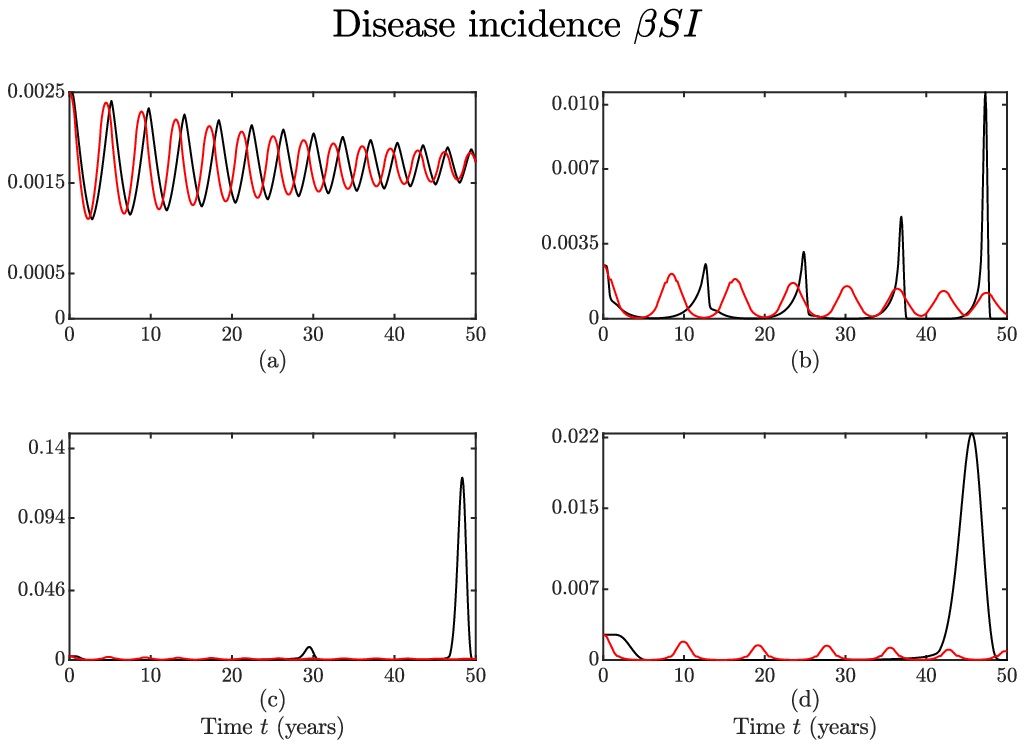}
    \caption{ Dynamics of the disease incidence $\beta S I$ as predicted by the QSSA model (\ref{TPB}) (red lines) and by the GSPT model (\ref{fast_ts}) (black lines). Panel (a):  $\psi=0.3,\,\phi=10$. Panel (b): $\psi=0.8,\,\phi=10$. Panel (c): $\psi=0.3,\,\phi=50$. Panel (d): $\psi=0.8,\,\phi=50$. Other parameter values and initial conditions as given in Table \ref{tab:param_table}.}
         \label{fig:dyn_epi_sus}
\end{figure}

In Figs. \ref{fig:vaccine}-\ref{fig:dyn_epi_sus} we display the numerical solutions of models (\ref{TPB}) and \eqref{fast_ts}, as well as the critical manifold (\ref{crit_manifC2}), 
for the four possible combinations of the factors $\psi$ and $\phi$ defining the imitation game parameters $\delta$ and $\theta$, as indicated in Table \ref{tab:param_table}. Namely: (a) $\psi=0.3,\,\phi=10$; (b) $\psi=0.8,\,\phi=10$; (c) $\psi=0.3,\,\phi=50$; (d) $\psi=0.8,\,\phi=50$.
Specifically, we combine cases in which: i) $\delta$ is considerably smaller than $p_c$ ($\psi=0.3$) or quite close to it ($\psi=0.8$); ii) $\theta$ is medium ($\phi=10 $) or relatively large ($\phi=50 $). Note that, in the case of absence of action enacted by the PHS ($ \gamma =0$), $\theta$ is the slope of the reactivity with respect to the information on the disease spread.

We report in Fig. \ref{fig:vaccine} the dynamics of the vaccine uptake of newborns, $p$; in Fig. \ref{fig:dyn_epi_inf} the dynamics of the fraction of infectious individuals normalized with respect to the endemic value in (\ref{EEcomp}), $I/I^*$; in Fig. \ref{fig:dyn_epi_sus} the dynamics of the disease incidence, $\beta S I$.

From Figs. \ref{fig:vaccine}-\ref{fig:dyn_epi_sus}, we note a generalized and very remarkable discrepancy between the solutions by the QSSA (model (\ref{TPB}), red lines) and those by the GSPT (model (\ref{fast_ts}), black lines). In any case, the solutions of both the models converge towards the endemic equilibrium. However, in the case (a), the damped oscillations of $p$ predicted by the QSSA have much larger  period than those predicted by the GSPT (Fig. \ref{fig:vaccine}a). In the cases (b)-(c)-(d), the orbits of the model \eqref{fast_ts} exhibit transitory oscillations by approaching $\{ p=1 \}$, while those of model  (\ref{TPB}) approach the endemic equilibrium after a short passage near $\{ p=1 \}$.

Further, in the case (a), the disease prevalence and incidence predicted by the
  model \eqref{fast_ts} resemble the shape of their QSSA counterpart (Fig. \ref{fig:dyn_epi_inf}a and Fig. \ref{fig:dyn_epi_sus}a). At variance, in the cases (b)-(c)-(d), the disease prevalence and incidence predicted by the GSPT  showcase effect of the delayed loss of stability of the system through a slow passage near $I=0$ and $SI = 0$, respectively.
From Fig. \ref{fig:dyn_epi_inf} we also note that for both the models varying $\psi$ and $\phi$ (and hence, $\delta$ and $\theta$) affects the timing of the  oscillations of the disease prevalence. In particular, increasing the values of $\psi$ and/or $\phi$ {increases the oscillation period}. 

A less pronounced but nonetheless important discrepancy is also observed in Fig. \ref{fig:vaccine} between the solutions of GSPT (model (\ref{fast_ts})) and the values of the critical manifold (\ref{crit_manifC2}) at the solutions of GSPT (black dashed lines). The discrepancy is evident in all the cases except for the case (a), where the orbits of both models (\ref{TPB}) and \eqref{fast_ts}  never approach $\{ p=1 \}$, and the critical manifold (\ref{crit_manifC2}) gives an excellent approximation of the behaviour of the model \eqref{fast_ts}. In the cases (b)-(c)-(d), the discrepancy is not limited to very short periods but it extends from some months (Fig. \ref{fig:vaccine}b) to many years (Fig. \ref{fig:vaccine}d), which can have a remarkable impact from the Public Health viewpoint.

 \FloatBarrier

\section{Reconsidering the case in which prevalence is the main information}\label{sec_app_C1}

In this section, we reconsider our initial imitation game equation (\ref{eqt_imit}) and 
briefly examine the case that the main information on the spread of the disease is the  \textit{prevalence} at time $\tau$, namely \begin{equation*}
M_d(\tau)=I(\tau).
\end{equation*}
This case was partially investigated in the paper \cite{DELLAMARCA2021} by using a QSSA approach. We now complete the analysis by using tools from GSPT. 

Note that we continue to assume that  $M_v$ is given by the vaccine uptake of newborns: $M_v(\tau)=p(\tau)$. 

The corresponding model in the fast time scale $t$ now reads:
\begin{equation}
\begin{aligned}
		S'&=\varepsilon(\mu(1-p)-\mu S-\beta SI),\\
		I'&=\varepsilon(\beta SI - (\nu+\mu) I),\\ 
		p'&=  p(1-p)(\delta + \theta_{\text{pre}}  I  - p  )+  \gamma (1-p),
\end{aligned}\label{SIRpfastC1}
\end{equation}
where 
$$\theta_{\text{pre}}=\dfrac{\theta_1}{\alpha_1}$$
 is the corresponding factor of perceived risk of infection, and $k$, $\delta$ and $\gamma$ as defined in (\ref{normaliz}).

We recall the expression of the endemic equilibrium $E^*=(S^*, I^*, p^*)$ of model \eqref{SIRpfastC1}, whose components are  given by
\begin{equation}
\label{EEcompC1}
    S^* = \dfrac{1}{\mathcal{R}_0}, \quad I^* = \dfrac{\mu }{\nu + \mu}\left( p_c - p^*\right),\quad
    p^* = \dfrac{\delta + \dfrac{\theta\mu }{\mu+\nu}p_c + \sqrt{\left(\delta +\dfrac{\theta\mu }{\mu+\nu}p_c\right)^2 + 4 \gamma \left(1 +\dfrac{\theta\mu }{\mu+\nu} \right)} }{2 \left(1 + \dfrac{\theta\mu }{\mu+\nu}\right)}.\end{equation}

The critical manifold is
\begin{equation}
\begin{aligned}\label{crit_manifC1}
    \mathcal{C}_0 =& \left\lbrace S\ge 0, I \ge 0, S+I \leq 1, p \in [0,1]\; |\; p(1-p)(\delta + \theta_{\text{pre}} I - p) + \gamma (1-p) = 0 \right\rbrace\\
    = & \{p = 1\} \cup \{p^2 - (\delta + \theta_{\text{pre}} I)p - \gamma  = 0\} =: A \cup B ,
\end{aligned}
\end{equation}
where $B$ is described by
\begin{equation}\label{xi}
    p = \xi(I) = \dfrac{\delta + \theta_{\text{pre}} I + \sqrt{(\delta + \theta_{\text{pre}} I)^2 + 4 \gamma}}{2 }.
\end{equation}

With the same argument employed in Section \ref{sec:GSPT_c2}, we obtain that  the Jacobian of the fast system \eqref{SIRpfastC1} with $\varepsilon = 0$ has two zero eigenvalues and a third eigenvalue given by
\begin{align*}
    \lambda 
&=  ( p^2 - (\delta + \theta_{\text{pre}} I)p - \gamma) + (1-p) (\delta + \theta_{\text{pre}} I - 2 p).
\end{align*}
On $B$, the eigenvalue is
\begin{equation*}
    \lambda = (1-p) (\delta + \theta_{\text{pre}} I - 2 p) < 0,
\end{equation*}
thus $B$ is always locally attractive when it exists, independently of the value of $\mathcal{R}_0$.

We now study the behaviour of the system when $p\approx 1$ (i.e. close to $A$). 
As in Section \ref{sec:GSPT_c2}, we have that the quantity
$S+I$ converges to zero.

The eigenvalue on $A$ is
\begin{equation*}
    \lambda = p^2 - (\delta + \theta_{\text{pre}} I)p - \gamma.
\end{equation*}
Thus,
\begin{equation*}
\begin{cases}
\lambda >0 &\text{if }\,  \dfrac{\delta + \theta_{\text{pre}} I + \sqrt{(\delta + \theta_{\text{pre}} I)^2 + 4 \gamma}}{2}<p\leq 1,\\
\lambda <0 &\text{if }      \, 0 \leq p < \dfrac{\delta + \theta_{\text{pre}} I + \sqrt{(\delta + \theta_{\text{pre}} I)^2 + 4 \gamma}}{2}.
\end{cases}
\end{equation*}
Since the corresponding eigenvalue $\lambda$ changes its sign, the region $A$ is attractive until the intersection with the curve (\ref{xi}), it then becomes repelling and, after a delay, the dynamics lands on the curve $B$. Depending on the value of $\mathcal{R}_0$ and $S_{\text{in}}$, where $S_{\text{in}}$ indicates the entrance of an orbit in a neighbourhood of $\{ p=1 \}$, we can observe two different behaviours:
\begin{itemize}
    \item Case I: if $\mathcal{R}_0 < 1$, or $\mathcal{R}_0 > 1$ and $S_{\text{in}}<1/\mathcal{R}_0$, then $I(\tau)$ is decreasing and the orbits converge to $B$ after a delay $T_E$ determined with the entry-exit function 
    $\int_0^{T_E}\lambda(\tau)\text{d}\tau=0$; hence the exit time $T_E$ is given implicitly by
    $$
    (1-\delta - \gamma )T_E= \theta_{\text{pre}}\int_0^{T_E}I(\tau)\text{d}\tau.
    $$
    A similar entry-exit phenomenon, with orbits eventually landing on a different branch of the critical manifold, was already observed  in the paper \cite{achterberg2022minimal}. The canard-like behaviour is confirmed applying the results in the paper \cite{krupa2001trans} (see also \cite{kaklamanos2022entry}). We note that, however, we have no explicit formula for $I(\tau)$ on $p=1$;
    \item Case II: if $\mathcal{R}_0 >1$ and $S_{\text{in}}>1/\mathcal{R}_0$, then $I(\tau)$  changes its monotonicity and the convergence towards $B$ only happens after a short excursion away from it. In this case, the known formulas for entry-exit functions cannot be applied.
\end{itemize}
\begin{figure}[t!]
  \centering
\includegraphics[width=0.8\textwidth]{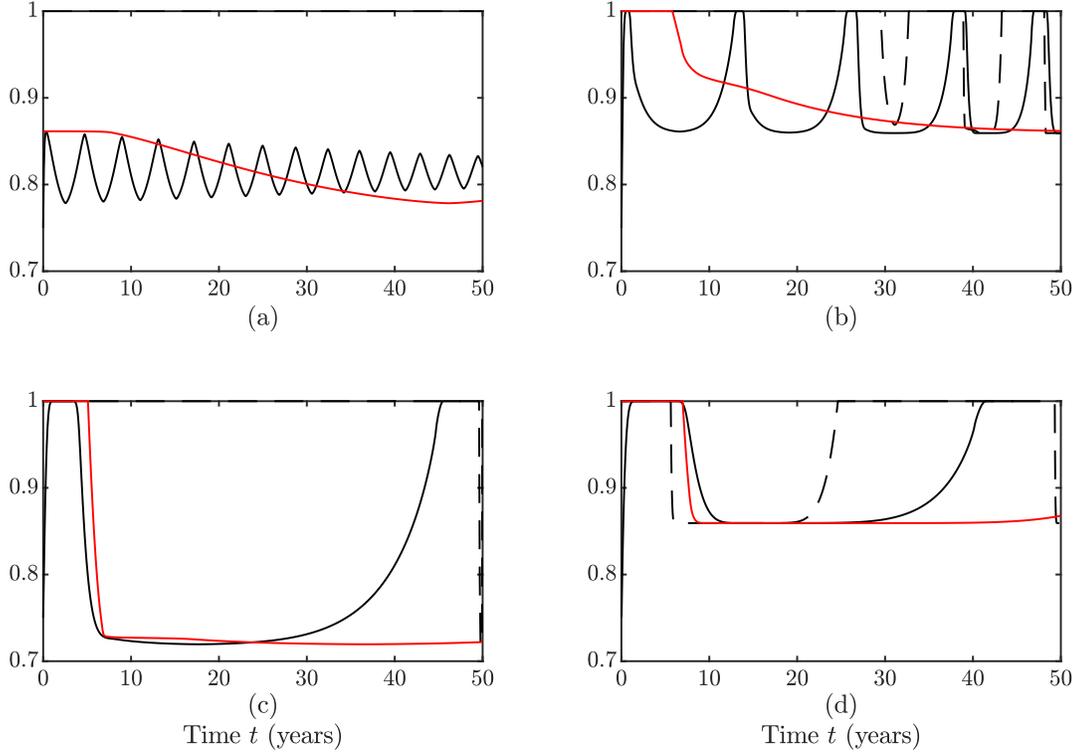}
    \caption{Information based on the disease prevalence: impact of the finiteness of the rate of switching strategy $k$. Vaccine uptake of newborns as predicted by  the QSSA model proposed in the paper \cite{DELLAMARCA2021} (red lines) and by the GSPT model (\ref{SIRpfastC1}) (black lines). The black dashed
lines represent the critical manifold \eqref{crit_manifC1}. Panel (a):  $\psi=0.3,\,\phi=10$ in (\ref{tetaref}). Panel (b): $\psi=0.8,\,\phi=10$ in (\ref{tetaref}). Panel (c): $\psi=0.3,\,\phi=50$ in (\ref{tetaref}). Panel (d): $\psi=0.8,\,\phi=50$ in (\ref{tetaref}). Other parameter values and initial conditions as given in Table \ref{tab:param_table}.}
         \label{fig:vaccineC1}
     \end{figure}
     
\begin{figure}[t!]
 \centering
\includegraphics[width=0.8\textwidth]{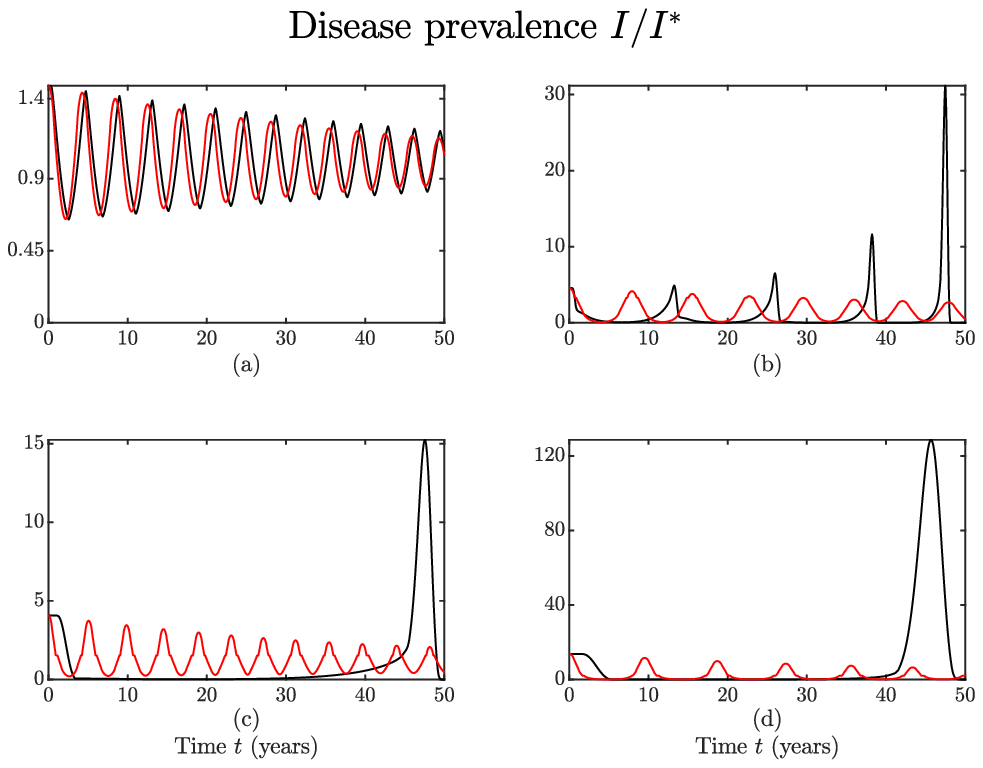}
    \caption{ Information based on the disease prevalence: dynamics  of the fraction of infectious individuals as predicted by the QSSA model proposed in the paper \cite{DELLAMARCA2021}
(red lines) and by the GSPT model (\ref{SIRpfastC1}) (black lines), and normalized with respect to the endemic equilibrium
value $I^*$ given in \eqref{EEcompC1}. Panel (a):  $\psi=0.3,\,\phi=10$  in (\ref{tetaref}). Panel (b): $\psi=0.8,\,\phi=10$  in (\ref{tetaref}). Panel (c): $\psi=0.3,\,\phi=50$  in (\ref{tetaref}). Panel (d): $\psi=0.8,\,\phi=50$  in (\ref{tetaref}). Other parameter values and initial conditions as given in Table \ref{tab:param_table}.}
         \label{fig:dyn_InfC1}
\end{figure}
In Figs. \ref{fig:vaccineC1}-\ref{fig:dyn_InfC1}, we report the numerical solutions of model (\ref{SIRpfastC1}), as well as the critical manifold (\ref{crit_manifC1}), and compare them with those by the QSSA approach. 
We display in Fig. \ref{fig:vaccineC1} the dynamics  of the vaccine uptake of newborns, $p$; in Fig. \ref{fig:dyn_InfC1} the dynamics of the fraction of infectious individuals normalized with respect to the endemic value in (\ref{EEcompC1}), $I/I^*$.
The parameters used are the same as in Table \ref{tab:param_table}, with the exception of the value of $\theta_{\text{pre}}$. Indeed, in this case we set
\begin{equation}
   \label{tetaref} 
\theta_{\text{pre}} = \phi 450, \quad \text{with } \phi =  \{10,50\}.
\end{equation}
From Fig. \ref{fig:vaccineC1}, we note that even in the scenario of prevalence-based information, the ``exact'' dynamics of $p$ (black lines) largely differs from the one predicted by the QSSA method (red lines). Moreover, the solutions of the GSPT model (\ref{SIRpfastC1}) may differ from the corresponding values along  the critical manifold \eqref{crit_manifC1} (black dashed lines) for time intervals of some years (Fig. \ref{fig:vaccineC1}b and Fig. \ref{fig:vaccineC1}d), which may have a remarkable impact from the Public Health viewpoint.

Similarly to what observed in Section \ref{sec:numerics}, from  Fig. \ref{fig:dyn_InfC1} we also note that for both the models increasing $\delta$ and/or $\theta_{\text{pre}}$ increases the oscillation period of the disease prevalence. 

\FloatBarrier

\section{Concluding remarks}\label{sec:concl}

The aims of this work were twofold. To start, we first considered  and studied the impact of the use of disease incidence instead of disease prevalence in the context of the vaccination imitation game. This assumption allows for a more realistic model because the incidence of a disease is a more widely diffuse datum than the prevalence. The study of the proposed model was done both for small and for very large parameter $k$, in the second case by using the QSSA. 

It is of interest to note that the QSSA provides a model of the spread and control of an SIR-like infectious disease that extends -- on a mechanistic ground  -- the purely phenomenological model \cite{domasa} to the important case where the information on the disease spread is the incidence, not the prevalence. As in the paper \cite{domasa}, also here oscillations do not occur. However, here we have a rationale: the lack of oscillations is related to the fact that the rate of strategy change is too large. The physical reason of this lack of limit cycles and other oscillating structures is that here we are in a regime of extremely volatile public opinion. As a consequence, there is no opinion-induced delay with respect to the information on the disease spread. Another important point to stress is that the QSSA  allows to precisely identify the role and the impact of the parameter $\gamma$ (which summarizes the efforts of PHS to convince hesitant people to change their vaccine-related decisions) on the vaccine uptake $p$, which turns out to be a \textit{nonlinear} function of both the disease incidence and of $\gamma$.

Then, we investigate the impact of the boundedness of the large parameter $k$. This allowed us to compare the analytical approaches of QSSA and GSPT.

The main result obtained by means of the GSPT approach lies in the significant extension of the QSSA approach, which is only valid in the limit as $\varepsilon \to 0$ (equivalently, $k \to +\infty$) that may be excessively sharp. As we remarked in Section \ref{sec:GSPT_c2}, instead, GSPT provides results for  $k$ large \emph{but} finite (equivalently $\varepsilon$ small \emph{but} strictly positive), which is more realistic from a behavioural switching point of view. This is crucial for the system under study, due to the presence of a transversal intersection of two parts of the critical manifolds. Orbits travelling close to this intersection do not always follow the stable branches of the critical manifold; rather, they spend a non-negligible amount of time in the vicinity of a repelling branch. This part of the dynamics can be characterized through the use of the so-called entry-exit function, as we showcased in Sections \ref{sec:GSPT_c2} and \ref{sec_app_C1}.

In the case analysed in Section \ref{sec_app_C1}, the canard-like behaviour can be explained through the application of known analytical results. The model we focus on for most of the work, i.e. the one in Section \ref{sec:GSPT_c2}, shares many similarities to the one in Section \ref{sec_app_C1}. However, to the best of the authors' knowledge, there is not general analytical result for the behaviour of multiple time scales system near a transcritical point in such a setting. A higher dimensional result in the spirit of the paper \cite{krupa2001trans} would be precious both from a purely theoretical point of view, and from its application potential.

Of course, this study suffers a number of limitations. We may mainly mention three: i) from the epidemiology viewpoint, the adopted model is a non-spatial and deterministic mean-field model; ii) from the behavioural viewpoint, the model is equipped with a relatively simplistic evolutionary model of opinion change; iii) we have no precise, universal way to quantify the range of validity for $k$ (equivalently, $\varepsilon$). Finally, we explicitly stress that our model is far more apt to describe the control of the spread of a childhood infectious disease than to deal with COVID-19, which would require a far more detailed model of both spread and control.

\bigskip

\noindent \textbf{Acknowledgements.}
The present work was performed under the auspices of the National Groups for Mathematical Physics (GNFM) (members: R.D.M. and A.d'O.) and for Mathematical Analysis, Probability and their Applications (GNAMPA) (member: S.S.) of the Italian 
National Institute for Advanced Mathematics (INdAM).

M.S. and S.S. were supported by the Italian Ministry for University and Research (MUR) through the PRIN 2020 project ``Integrated Mathematical Approaches to Socio-Epidemiological Dynamics'' (No. 2020JLWP23).

\appendix
\section{Appendix}

\subsection{Global stability of $E^P$ for model (\ref{SIRpcompleto})}\label{App1}
Let us assume that $\gamma>\bar\gamma$, where $\bar\gamma=1-\delta$. From the differential equation (\ref{dotpC2}) it immediately follows that
$$\dot p\geq k( p(\delta-1)+\gamma )(1-p),$$
implying that
$$\liminf_{t\to+\infty} p\geq \min\left(1,\dfrac{\gamma}{1 -\delta}\right).$$
Since $\gamma>\bar\gamma$, then $\liminf_{t\to+\infty} p=1$, thereby showing that $E^P$ is GAS. 

Conversely, if $\gamma<\bar\gamma$, then the instability of $E^P$ easily follows by linearising the differential equation (\ref{dotpC2}) at $E^P$.

\subsection{Global stability of $E^0$ for model (\ref{SIRpcompleto})}\label{App2}
Assume now that $\gamma_c<\gamma<\bar\gamma$, where $\gamma_c=p_c( p_c - \delta)$. Consider the following inequality
$$\dot p \geq kp(1-p)\left(\delta- p+\dfrac{\gamma}{p}\right),$$
it follows that
$$\liminf_{t\to+\infty} p=p^0.$$
Moreover, the above minimum limit implies that for large times
$$\dot S\leq \mu(1-p^0 -S)$$
and in turn that
$$\dot I\leq \beta I(p_c-p^0).$$
Simple algebra shows that $\gamma>\gamma_c$ is equivalent to $p^0>p_c$, and the global stability of $E^0$ immediately follows.

Conversely, if $\gamma<\gamma_c$ (i.e. $p^0<p_c$), then the instability of $E^0$ follows from the linearized equation for the infectious fraction: $\dot i=\beta i (p_c-p^0).$

\subsection{Global stability of $E^0$ for model (\ref{TPB})}\label{App3}
By defining $\sigma=S+I$, from system (\ref{TPB}) we obtain
$$\dot \sigma=\mu(1-\zeta(S,I))-\mu\sigma-\nu I\leq \mu (1-p^0)-\mu\sigma,$$
it follows that
$$\limsup_{t\to+\infty} \sigma=1-p_0.$$
The above maximum limit implies that for large times
$$\dot I\leq \beta I(p_c-p^0).$$
Hence, if $p^0>p_c$, then   $E^0$ is globally attractive.

\subsection{Global stability of $E^*$ for model (\ref{TPB})}\label{App4}
The Jacobian matrix of system (\ref{TPB}) evaluated at the equilibrium $E^*$ reads
$$J=\left(\begin{array}{cc}
     -\mu \dfrac{\partial \zeta}{\partial S}\left(\dfrac{1}{\mathcal{R}_0},I^*\right) -\beta I^*-\mu &   -\mu \dfrac{\partial \zeta}{\partial I}\left(\dfrac{1}{\mathcal{R}_0},I^*\right) -(\nu+\mu)   \\
     \beta I^* & -(\nu+\mu) 
\end{array}\right).$$
Since tr$ J<0$ and det$J>0$, the eigenvalues of $J$ have negative real parts.
Further, in $\Omega^*$ there are no closed orbits since
$$\text{div}\left(\dfrac{1}{I}(\dot S, \dot I)\right)=-\dfrac{\mu}{I}   \dfrac{\partial \zeta}{\partial S}-\beta -\dfrac{\mu}{I}<0.$$
Thus, by the Poincar\'e-Bendixon thricotomy  it follows that  $E^*$ is GAS in $\Omega^*$.

\subsection{Change of variables}\label{change_var}
In order  to reduce the numerical stiffness of the model  \eqref{fast_ts},  we employ a change of the state variables of the model. Specifically, we introduce:
\begin{equation*}
    x:= \ln(S), \quad y:= \ln(I), \quad z:= \ln\left(\dfrac{p}{1-p}\right),
\end{equation*}
i.e.
\begin{equation*}
    S= e^x, \quad I = e^y, \quad p= \dfrac{e^z}{1+e^z}.
\end{equation*}
By deriving the new variables with respect to the  time $t$, we obtain 
\begin{align*}
    x' =& \; \varepsilon \left(\mu \dfrac{e^{-x}}{1+e^z} - \beta e^y - \mu\right),\\
    y' = & \; \varepsilon \left(\beta e^x - \nu -\mu \right),\\
    z' = &\; \delta + \theta \beta e^{x+y} -  \dfrac{e^z}{1+e^z} + \gamma \dfrac{1+e^z}{e^z}.
\end{align*}
The obtained system is remarkably less numerically stiff than the original one.

\FloatBarrier

\bibliographystyle{abbrv}
\bibliography{biblio}

\end{document}